\documentclass[aps,prb,twocolumn,floatfix,superscriptaddress,longbibliography]{revtex4-2}

\usepackage{amsmath}
\usepackage{amssymb}
\usepackage{times}
\usepackage{braket}
\usepackage[pdftex]{graphicx}
\usepackage{blindtext}
\usepackage{color}

\renewcommand{\vec}[1]{\boldsymbol{#1}}

\newcommand{\change}[1]{\textcolor{black}{#1}}
\newcommand{\changeb}[1]{\textcolor{black}{#1}}
 
\begin{document}

\title{Quaternary-digital data storage based on magnetic bubbles in anisotropic materials}

\author{B{\"o}rge G{\"o}bel}
\email[Corresponding author. ]{boerge.goebel@physik.uni-halle.de}
\affiliation{Institut f\"ur Physik, Martin-Luther-Universit\"at Halle-Wittenberg, D-06099 Halle (Saale), Germany}

\author{Ingrid Mertig}
\affiliation{Institut f\"ur Physik, Martin-Luther-Universit\"at Halle-Wittenberg, D-06099 Halle (Saale), Germany}

\date{\today}

\begin{abstract}
The topologically non-trivial nano-whirls, called magnetic skyrmions, are often considered attractive for spintronic applications like the racetrack data storage device. However, skyrmions do not move parallel to applied currents and typically do not coexist with other nano-objects, making the realization of such storage concepts difficult. Herein we consider materials with an anisotropic DMI, like tetragonal Heusler materials, and show that four distinct types of topologically trivial bubbles can coexist. We show that each of them can be written by spin torques and that they can be distinguished using a single magnetic tunneling junction. Due to their trivial topology, the four types of bubbles move parallel to applied electrical currents and remain equidistant under motion. Still, the bubbles have a small size and high stability comparable to topologically non-trivial spin textures. This allows to construct a quaternary-digit-based racetrack storage device that overcomes the two mentioned drawbacks of skyrmion-based racetracks.
\end{abstract}


\maketitle

\section{Introduction}
Over the past decades, the concept of topology has been explored in several physical systems like for the quantum Hall effect~\cite{thouless1982quantized} or in topological insulators~\cite{kane2005z}. In magnetism, non-collinear spin textures, like skyrmions, can be characterized by a topological invariant as well, $N_\mathrm{Sk}=\frac{1}{4\pi}\int\vec{m}\cdot(\partial_{x}\vec{m}\times \partial_{y}\vec{m})\,\mathrm{d}^2r$.
Magnetic skyrmions have been observed in 2009 in the B20 material MnSi \cite{muhlbauer2009skyrmion} in which the broken inversion symmetry gives rise to a Dzyaloshinskii-Moriya interaction~\cite{dzyaloshinsky1958thermodynamic,moriya1960anisotropic} (DMI) that fixes the chirality of the skyrmions. 

The topological character of such spin textures is often considered to be responsible for their stability: In an infinitely extended continuum, a skyrmion could never be transformed into a topologically trivial phase like a ferromagnet or a helical phase \cite{nagaosa2013topological}. However, in reality they exist on a crystal lattice, which is why this so-called topological protection does not exist anymore: Magnetic skyrmions can be generated~\change{\cite{everschor2017skyrmion}}, annihilated~\cite{romming2013writing}, \changeb{transformed~\cite{zhang2015magnetic}} and even fractional skyrmions~\cite{parkin2021observation} as well as topologically trivial bubbles~\cite{moon2015magnetic,jena2020elliptical,peng2020controlled} have been observed.

\begin{figure*}[t!]
  \centering
  \includegraphics[width=\textwidth]{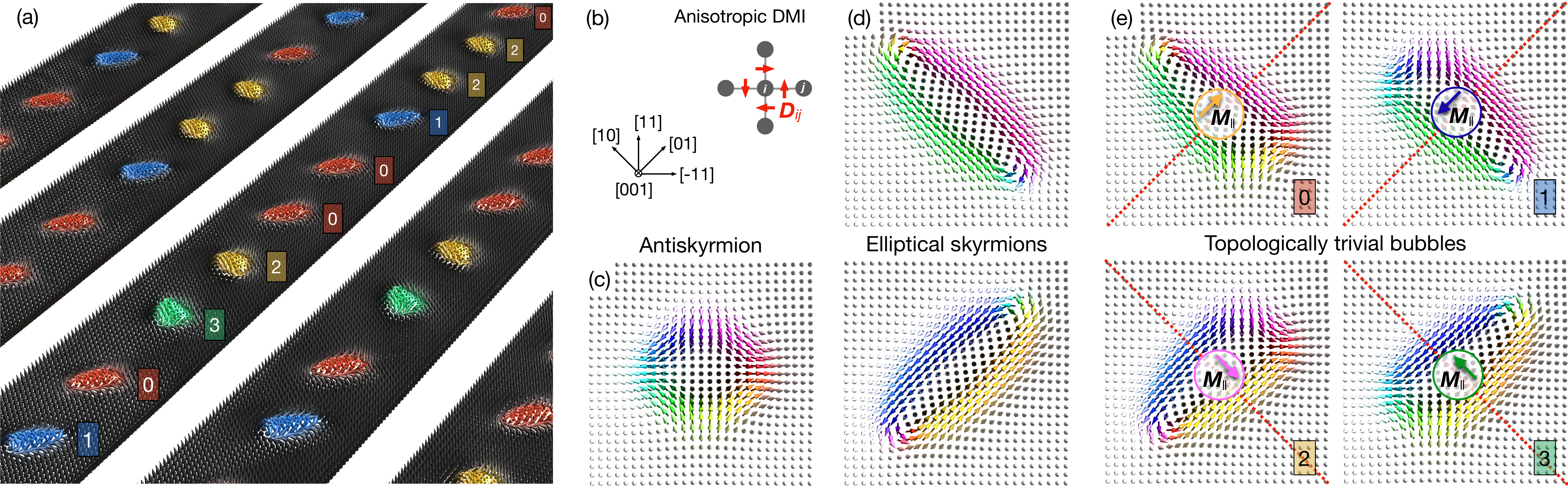}
  \caption{Distinct spin textures in materials with anisotropic DMI. (a) Overview of the suggested bubble-based racetrack built from Heusler materials. The color encodes the four different types of bubbles as quaternary digits `0'--`3'. (b) The DMI vectors $\vec{D}_{ij}$, corresponding to an anisotropic DMI as in the Heusler material Mn$_{1.4}$Pt$_{0.9}$Pd$_{0.1}$Sn, and the high-symmetry crystallographic directions are shown. (c) Stabilized antiskyrmion and (d) both types of elliptical skyrmions, respectively. (e) Four different types of topologically trivial bubbles. In all cases, these bubbles are the combination of a skyrmion-like and an antiskyrmion-like part, separated by the dashed line. The topological charge is zero. The in-plane net magnetization $\vec{M}_\parallel$ is indicated. All snapshots show a part of a circular nano disk with a diameter of $500\,\mathrm{nm}$. The size of the objects is $\sim 200\,\mathrm{nm}$ similar to the experimental findings~\cite{jena2020elliptical,peng2020controlled}.}
  \label{fig:anisotropic}
\end{figure*}

Bubbles are non-collinear objects like skyrmions, however, they are typically stabilized by the achiral dipole-dipole interaction in centrosymmetric materials, in which the DMI is forbidden by symmetry. They can consist of parts that resemble skyrmions and parts that resemble antiskyrmions, which are the antiparticles of skyrmions with an opposite topological charge. If these two parts come in equal proportions, a bubble is topologically trivial, i.\,e. $N_\mathrm{Sk}=0$. The increased size and flexibility in shape and orientation of bubbles in centrosymmetric materials lead to lower storage densities of bubble-based shift devices~\cite{bobeck1969magnetic,michaelis1975magnetic} compared to the skyrmion-based racetrack storage device~\cite{parkin2008magnetic}. 
Both concepts operate without moving parts and encode data via the presence or absence of skyrmions or bubbles at predefined positions, respectively. However, the skyrmion-based racetrack \cite{parkin2004shiftable,parkin2008magnetic, parkin2015memory,sampaio2013nucleation, fert2013skyrmions} is hampered by the skyrmion Hall effect: Due to the topological charge of magnetic skyrmions, they do not move parallel to an applied current but are instead pushed towards the edge of the racetrack \cite{iwasaki2013current,jiang2017direct,litzius2017skyrmion,gobel2018overcoming}. This severe problem of skyrmions could potentially be solved by considering alternative magnetic nano-object \cite{gobel2020beyond}.

Herein, we show that topologically trivial bubbles in materials with an anisotorpic DMI combine the advantages of chiral skyrmions and achiral bubbles and allow to construct a high-performance bubble-based racetrack data storage device. We use micromagnetic simulations to stabilize topologically trivial bubbles and show that their skyrmion part is always oriented along one of the four \{10\} directions (this nomenclature is an abbreviation for the [100], [$\overline{1}$00], [010] and [0$\overline{1}$0] directions). This eliminates the unfavorable deformability of bubbles in achiral materials. It allows us to construct a quaternary-digit-based data storage device; cf. Fig. \ref{fig:anisotropic}(a). The four distinct types of bubbles are highly stable and, due to the vanishing topological charge, these bubbles can be driven parallel to an  applied  current along the racetrack (Fig. \ref{fig:motion}).  We  show that the four bubbles can be generated reliably by spin-currents that are controlled by only two voltages and we show that they can be detected and distinguished by a magnetic tunneling junction (MTJ, Fig. \ref{fig:generation}). Indeed, oriented topologically trivial bubbles have recently been observed in the Heusler material Mn$_{1.4}$Pt$_{0.9}$Pd$_{0.1}$Sn which has a D$_{2d}$ symmetry \cite{jena2020elliptical,peng2020controlled} and is characterized by an anisotropic DMI.

\section{Results and Discussion}
\subsection{Spin textures in isotropic chiral magnets}
To understand how topologically trivial bubbles excel in material\changeb{s} with an anisotropic DMI, first we present the properties of such objects in materials with an isotropic DMI. An example is the B20 material MnSi~\cite{muhlbauer2009skyrmion}. In this material, the inversion symmetry is intrinsically broken, so that a so-called bulk DMI arises. This interaction favors Bloch skyrmions of either right- or left-handed helicity, corresponding to a helicitiy of $\pi/2$ and $-\pi/2$, respectively, depending on the sign of the DMI constant \cite{bogdanov2002magnetic}.  Also, as in all magnetic materials, dipolar interactions are present that we consider in the form of magnetostatic fields. These interactions favor the stabilization of skyrmions of both Bloch helicities $\pm \pi/2$ equally and therefore modify the magnetization profile of the DMI-dominated skyrmions in these materials only slightly.

Despite the energetic preference of skyrmions over all other non-collinear nano-objects, it is possible that other spin textures are metastable as well, at least theoretically. Higher-order skyrmions \cite{rozsa2017formation}, antiskyrmions \cite{ritzmann2018trochoidal} and topologically trivial bubbles \cite{sisodia2020skyrmion} have been metastabilized in materials with an isotropic DMI. Considering the latter object in the sense of a combination of half a skyrmion and half an antiskyrmion, it becomes apparent that the antiskyrmion part is strongly disfavored by the magnetic interactions and can only remain stable due to the remnant `topological protection' on the lattice: When we start from a circular object in a nanodisk and relax it to the nearest local minimum of the energy, the antiskyrmion part shrinks and becomes much smaller than the skyrmion part (Supplementary Figure 1(a) \cite{SupplementalMaterial}), similar to the simulations shown in~\cite{sisodia2020skyrmion}.
We have modeled this magnetic texture in the micromagnetic framework mumax3 \cite{vansteenkiste2011mumax,vansteenkiste2014design} according to the Landau-Lifshitz-Gilbert equation (see Supplementary Information for details \cite{SupplementalMaterial}).
%


Since the DMI and the dipole-dipole interactions are isotropic, the topologically trivial bubble can be rotated freely. In Supplementary Fig. 1 this has been realized using a magnetic field whose inplane component is continuously rotated (for details see \cite{SupplementalMaterial}). The net magnetization of the bubble $\vec{M}_\parallel$ remains aligned with $\vec{B}_\parallel$. The possibility to continuously reorient bubbles in materials with an isotropic DMI may be usable for spintronic applications like a nano-rotor. 
However, this flexibility renders them unfavorable for racetrack applications where a stability of the bits is essential. 

\begin{figure*}[t!]
  \centering
  \includegraphics[width=0.8\textwidth]{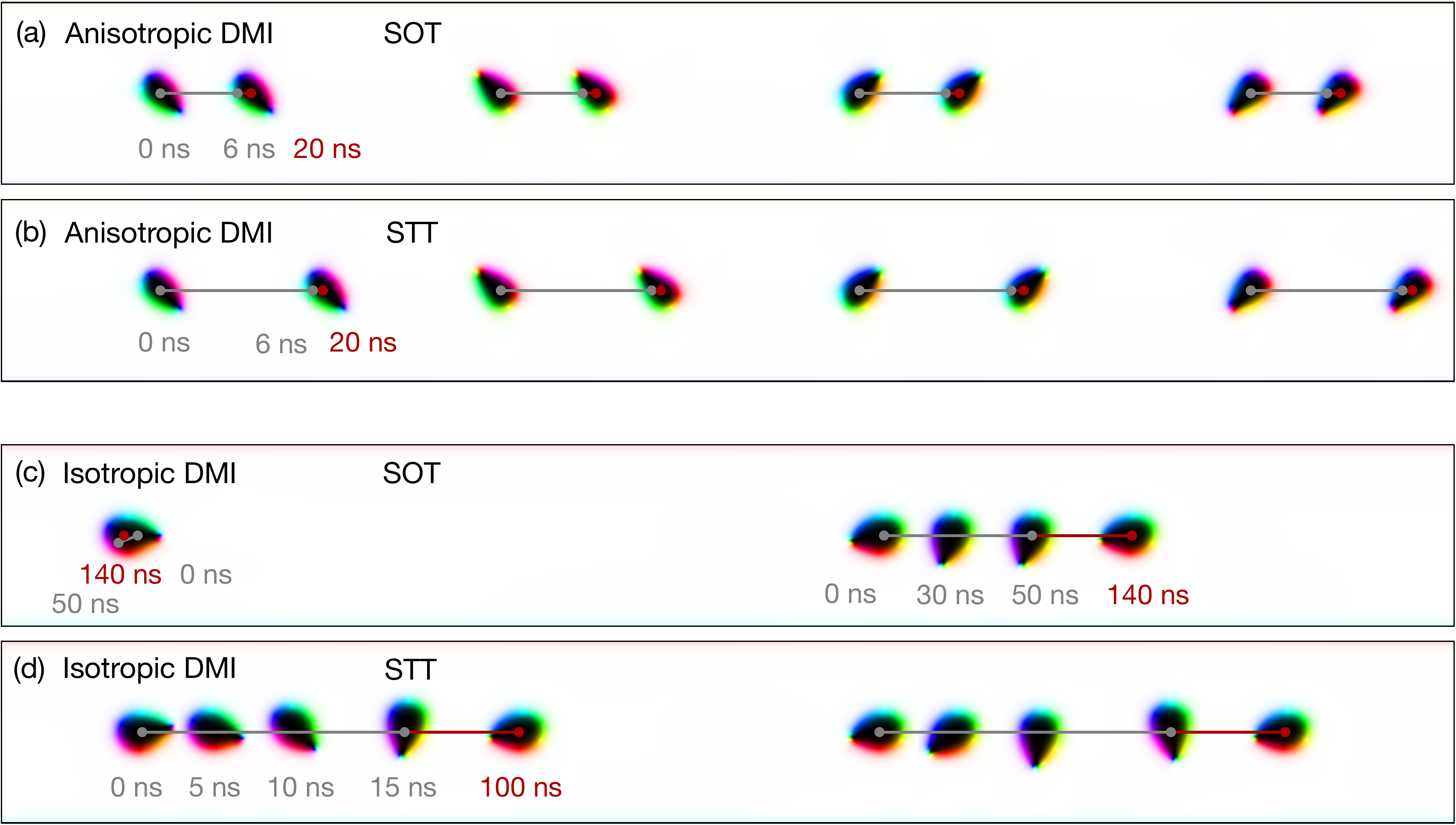}
  \caption{Current-driven motion of bubbles in systems with anisotropic and isotropic DMI. (a,b) Current-driven motion of the four different types of bubbles in a Heusler material with anisotropic DMI in the spin-orbit torque (SOT) and the spin-transfer torque (STT) geometries, respectively. The image is a composite of multiple snapshots. The line indicates how one bubble has moved. The current is applied for the first $6\,\mathrm{ns}$ (gray) and the system is relaxed for another $14\,\mathrm{ns}$ (red), afterwards. (c,d) The same for a B20 material with an isotropic DMI. (c) Only one of the two starting configurations moves in the SOT scenario. The current is applied for $50\,\mathrm{ns}$. (d) In the STT scenario, both bubbles move but the left bubble switches its orientation. The current was applied for $15\,\mathrm{ns}$. In all panels the (spin) current density is $30\,\frac{\mathrm{MA}}{\mathrm{cm}^2}$.}
  \label{fig:motion}
\end{figure*}

\subsection{Anisotropic topologically trivial bubbles in materials with an anisotropic DMI}

Next, we consider topologically trivial bubbles in magnets with an anisotropic DMI, like in the Heusler material Mn$_{1.4}$Pt$_{0.9}$Pd$_{0.1}$Sn. The anisotropic DMI innately stabilizes antiskyrmions. Due to the anisotropic orientation of the DMI vectors [Fig.~\ref{fig:anisotropic}(b)], antiskyrmions are not continuously rotational symmetric, in contrast to skyrmions. The antiskyrmion's Bloch parts (toroidal configuration) are always oriented along the \{10\} directions and the N\'{e}el parts (radial configuration) are oriented along the \{11\} directions (this means [110], [$\overline{1}$10], [1$\overline{1}$0] and [$\overline{1}\overline{1}$0]). 

Furthermore, as in every magnetic material, dipole-dipole interactions are present that favor the stabilization of Bloch skyrmions. For this reason, in this Heusler material, two types of elliptical Bloch skyrmions have also been observed \cite{jena2020elliptical,peng2020controlled}. From the simulations in Figs. \ref{fig:anisotropic}(d) it becomes apparent that these objects elongate along the \{10\} directions, depending on their handedness. This is because the anisotropic DMI energetically favors the clockwise Bloch parts along the [010] and [0$\overline{1}$0] directions and favors the counter-clockwise Bloch parts along the [100] and [$\overline{1}$00] directions. Therefore, these parts elongate which leads to an elongation of the whole skyrmion along the [100] direction for clock-wise Bloch skyrmions with helicity $-\pi/2$ and an elongation along the [010] direction for counter-clock-wise Bloch skyrmions with helicity $+\pi/2$. 

When we stabilize bubbles, again, we start from a half-skyrmion and a half-antiskyrmion in a nano-disk and relax the texture. Upon relaxation to the nearest energy minimum, the profile deforms; the skyrmionic part becomes elongated similar to the elliptical skyrmions discussed above. This is in contrast to B20 materials, in which the part that is unfavored by the DMI (in B20 the antiskyrmionic part) shrinks drastically. Furthermore, here, in the material with an anisotropic DMI,  both Bloch helicities lead to stable bubbles and the orientation of the bubbles is fixed. It is determined by the fixed orientation of the skyrmion part and corresponds to the elongation directions of the elliptical skyrmions. This means four distinct, stable configurations exist; they are shown in Figs. \ref{fig:anisotropic}(e). The clockwise Bloch parts are elongated along the [100] or [$\overline{1}$00] directions (labeled `0' and `1' bubble) and the counter-clockwise Bloch parts are oriented along the [010] or [0$\overline{1}$0] direction (labeled `2' and `3').

Due to the anisotropic DMI, these configurations are not easily transformable into each other. Applying an in-plane magnetic field, as was done for bubbles in B20 materials before (cf. Supplementary figure 1 \cite{SupplementalMaterial}), does not allow for a rotation of the bubbles in this case. The different configurations are separated by considerable energy barriers. Even tilting the field by $20^\circ$ instead of $1^\circ$, that was used in the case of bubbles in isotropic materials, leaves the initial bubble unchanged. For larger tilting angles, instead of rotating, the bubble transforms to an antiskyrmion, because for the considered parameters it has the lower energy \change{(Bubble in the nanodisk: $E=-4.272\,\mathrm{pJ}$ vs. antiskyrmion $E=-4.369\,\mathrm{pJ}$). Still a bubble has a lower energy than a skyrmion ($E=-4.181\,\mathrm{pJ}$) meaning that it has a comparable or even higher stability than the topologically non-trivial spin textures.}

In the recent experimental studies on coexisting antiskyrmions and skyrmions in Heusler materials, topologically trivial bubbles have been observed as well, and indeed the skyrmionic part is always oriented along one of the the four \{10\} directions: In the Supplementary Figure 8 of Ref. \cite{jena2020elliptical} and in Figure 2 of Ref. \cite{peng2020controlled} it was shown, that 4 different types of trivial bubbles can be generated by providing an in-plane magnetic field along several directions.

\subsection{Bubble-based racetrack storage device}

The \change{stability} of the four different types of bubbles in Heusler materials \change{is comparable to that of the topologically non-trivial spin textures which} allows to construct an improved data storage device. These bubbles constitute the quaternary digits `0', `1', `2' and `3' in the digital data storage. As we will show in the following, this concept does not only have double the storage density but it can also overcome both main disadvantages of a skyrmion-based racetrack in a system with an isotropic DMI: (i) The non-trivial topology of skyrmions leads to the emergence of a skyrmion Hall effect, i.\,e. a transverse deflection, when driven by currents \cite{iwasaki2013current,jiang2017direct,litzius2017skyrmion,gobel2018overcoming}. (ii) Data is encoded via the presence or absence of skyrmions at predefined positions but a sequence of skyrmions and missing skyrmions may relax to a periodic sequence of equidistant skyrmions due to attractive and repulsive interactions between them. Consequently, the information is lost.

As recently reported, Heusler materials can solve both problems individually: (i) If cut under the correct angle \cite{jena2020formation}, antiskyrmions driven by spin-orbit torques do not experience a skyrmion Hall effect \cite{huang2017stabilization}. (ii) Antiskyrmions and elliptical skyrmions can coexist in these samples \cite{jena2020elliptical,peng2020controlled,jena2020formation, sivakumar2020topological}. By using topologically trivial bubbles in these materials, we can combine both advantages: (i) Due to the absence of topological charge, $N_\mathrm{Sk}=0$, all bubbles move parallel to the current without any skyrmion Hall effect. (ii) The four different configurations coexist and remain stable even when driven by currents.

In Figs. \ref{fig:motion}(a,b) we show the motion of all different types of bubbles under spin-orbit torque (SOT) and spin-transfer torque (STT). \change{Both scenarios rely on the collective reorientation of the bubbles' magnetic moments due to the interaction with a spin (polarized) current. For the SOT~\cite{slonczewski1996current} the spin current is injected from the perpendicular direction and it is generated from a charge current in a heavy metal by the spin Hall effect. For the STT~\cite{zhang2004roles}, a charge current flows directly in the magnetic Heusler material and gets spin polarized. More simulation details are explained in the Supplementary Material \cite{SupplementalMaterial}.} In both scenarios, all four types of bubbles are rigidly shifted along the current direction, i.\,e. from left to right along the track, with the same velocity. This behavior is optimal for the reliability and performance of this device. 

For bubbles in B20 materials the dynamics is drastically different. As we have explained before, in a circular nanodisk, these bubbles can be rotated easily. Since the geometry of the racetrack is not rotational symmetric anymore, the bubbles in this material are aligned always along the racetrack (two opposite orientations). Even though these configurations are energetically favored, the energy barrier between them is much smaller than for the anisotropic case which brings about severe problems that prevent the construction of a bubble-based racetrack device in materials with an isotropic DMI: In the simulation under SOT in Fig. \ref{fig:motion}(c) it becomes apparent that the distance between the two bubbles changes because only one of the two bubbles reorients and moves. 
In the simulation under STT in Fig. \ref{fig:motion}(d) it is shown that the bits can flip because the two bubble configurations are only separated by such a small energy barrier. Therefore, both possible operation modes in materials with an isotropic DMI are unserviceable due to an inevitable loss of data.

\subsection{Generation and detection}

\begin{figure}[t!]
  \centering
  \includegraphics[width=\columnwidth]{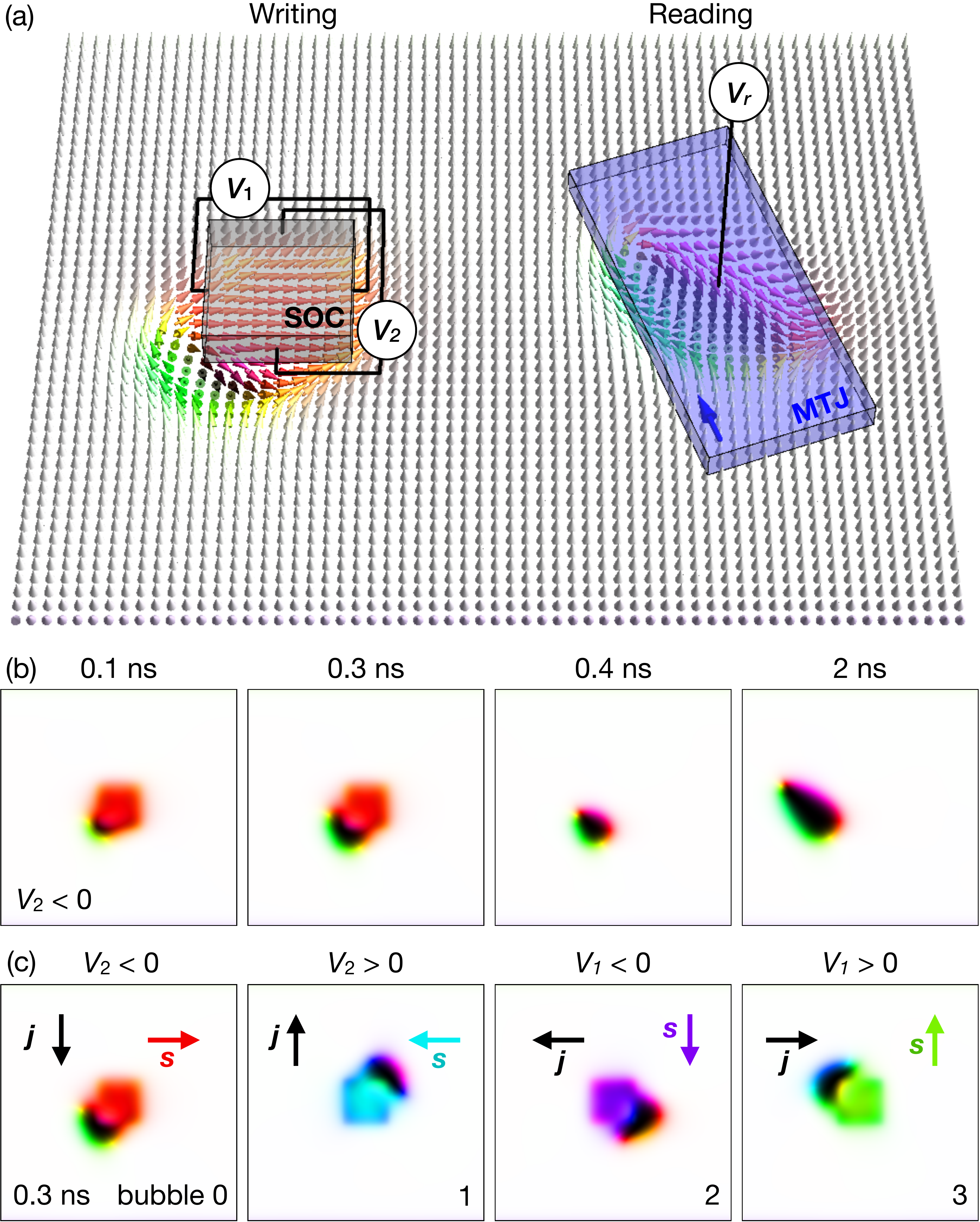}
  \caption{Generation and detection of topologically trivial bubbles in systems with isotropic and anisotropic DMI. (a) Left: Concept for writing the four different types of bubbles by spin currents. The arrows show the simulated texture after applying a current ($j_y\theta_{\mathrm{SH}}=-5\times 10^{14} \frac{\mathrm{A}}{\mathrm{m}^2}$) for $0.3\,\mathrm{ns}$. This current is generated by a negative $V_2<0$ and $V_1=0$ and leads to an injection of spins $\vec{s}\parallel\vec{x}\parallel [\overline{1}10]$ into the ferromagnet. Right: Concept for detecting and distinguishing the four types of bubbles by a magnetic tunneling junction (MTJ). (b) Sequence of snapshots upon creating a bubble `0'; cf. Fig. \ref{fig:anisotropic}(d). The current is applied for $0.3\,\mathrm{ns}$, i.\,e., in the first two panels. In the last two panels, the current is not applied anymore and the bubble relaxes. (c) The same procedure but for different current and spin orientations, generated by different applied voltages, as indicated. All four types of bubbles are creatable in a controlled manner.}
  \label{fig:generation}
\end{figure}

Before we conclude, we want to address the two other important ingredients that are needed to operate this bubble-based racetrack device: writing and detecting the carriers of information. 

For a controlled writing of the four different bubbles, we utilize a mechanism that is again unique for materials with an anisotropic DMI. We reorient a region of the ferromagnetic sample to point in-plane. This could be achieved by STT via currents from the perpendicular direction or via SOTs. We use the latter case because this allows to precisely control the in-plane orientation of the spins by two voltages $V_1$ and $V_2$, as shown in Fig. \ref{fig:generation}(a). 

By switching the magnetization locally in the plane, the magnetization relaxes at the boundary of this region according to the magnetic interactions. The anisotropic DMI leads to the formation of non-collinearities that constitute the rough profile of a topologically trivial bubble. In Fig. \ref{fig:generation}(b) the formation process under the applied spin current (first two panels) and the relaxation process (last two panels) are shown for spins that are oriented along [$\overline{1}1$]. These spins have been generated by $V_1=0$ and $V_2<0$ and give rise to bubble `0'. In Fig. \ref{fig:generation}(c) it is shown that the other types of topologically trivial bubbles can be generated by applying positive and negative voltages $V_1$ or $V_2$, respectively. Due to the anisotropic nature of the DMI, the direction of the in-plane magnetization in the square-shaped region uniquely determines the type of generated bubble.

For reading the bubble, we use their distinct net magnetizations: Bubble `0' has a net magnetization along the [01] direction, bubble `1' along [0$\overline{1}$], bubble `2' along [$\overline{1}$0], and bubble `3' along [10] [Fig.~\ref{fig:anisotropic}(e)]. This magnetization component $\vec{M}_\parallel$ can for example be detected by a magnetic tunneling junction; cf. Fig. \ref{fig:generation}(a). It is particularly elegant to use an MTJ with a fixed layer that is magnetized along an in-plane direction $\vec{M}_\mathrm{MTJ}$ that is not along the family of \{10\} or \{11\} directions, so that the projection of $\vec{M}_\parallel$ on $\vec{M}_\mathrm{MTJ}$ is different for all cases. For example, for the configuration shown in Fig. \ref{fig:generation}(a), the reading signal is positive but distinct for the bubbles `0' and `3' and negative but distinct for bubbles`1' and `2'. Using this method, a single MTJ can detect and distinguish all four distinct carriers of information.

\section{Conclusion}
In summary, magnetic bubbles in materials with an anisotropic DMI behave fundamentally different compared to bubbles in isotropic or centrosymmetric materials. Even though they are topologically trivial, they exist in 4 very stable and distinct configurations, based on which we have predicted a quaternary-digit-based racetrack data storage device. This concept doubles the storage capacity compared to binary-digit-based racetracks and solves the two critical drawbacks that, so far, have hindered the realization of a skyrmion-based racetrack storage: (i) The bubbles move parallel to the current without skyrmion Hall effect. (ii) Information is not encoded by the presence or absence of one type of non-collinear spin texture but by four distinct objects (two would already be sufficient but less efficient). This allows for a reliable device for which the sequence of information does not have to be equidistant.

\begin{acknowledgments}
We acknowledge valuable discussions with Stuart S. P. Parkin and Jagannath Jena about non-collinear spin textures in Heusler materials. This work is supported by SFB TRR 227 of Deutsche Forschungsgemeinschaft (DFG).
\end{acknowledgments}

\bibliography{short,MyLibrary}

\end{document}